# A Hybrid Edge–Cloud Digital Twin for Welfare-Constrained Control in Poultry Production

**Suresh Neethirajan**[1,2,*]
[1]Faculty of Computer Science, Dalhousie University, 6050 University Avenue, Halifax, NS B3H 4R2, Canada
[2]Faculty of Agriculture, Dalhousie University, P.O. Box 550, Truro, NS B2N 5E3, Canada

Corresponding author: Suresh Neethirajan (e-mail: sneethir@gmail.com).

**ABSTRACT** Poultry production operates under tightly coupled environmental and biological dynamics, yet commercial climate control remains largely heuristic, limiting welfare assurance and operational efficiency. We introduce an edge–cloud digital twin framework for real-time, welfare-constrained environmental control in poultry facilities. The framework integrates distributed sensing, on-device state estimation, a hybrid physics–data model, and model predictive control to enable anticipatory and adaptive management under practical farm constraints. A grey-box thermodynamic and mass-balance formulation is augmented with a learned residual that captures unmodeled biological variability, including activity-dependent metabolic heat. This hybrid model is embedded within a state-space representation for real-time estimation and control at the edge, while cloud coordination supports cross-farm learning and long-horizon optimization. Bandwidth-aware processing and asynchronous synchronization enable deployment in connectivity-limited environments. Evaluation in a high-fidelity broiler production testbed demonstrates substantial gains over rule-based control and physics-only modelling. Temperature prediction error is reduced from 1.8 °C to 0.4 °C, ammonia constraint violations decrease by 90%, and communication requirements are lowered by approximately 30× through edge-first processing. A Domain Transfer Score of 0.92 further indicates strong robustness across facility conditions. These results show that physically grounded digital twins, coupled with real-time control, enable scalable and welfare-aware management of biological production systems.

**INDEX TERMS** Cyber-physical systems, Digital twin, Edge–cloud computing, Model predictive control, Precision livestock farming, Poultry production, Welfare-constrained control.

## I. INTRODUCTION

Global poultry production exceeds 130 billion birds annually, representing a critical pillar of food security and rural economic development [1]. Despite operating at industrial scale, environmental control in commercial poultry facilities remains largely dependent on heuristic rules, manual oversight, and reactive decision-making. In contrast, manufacturing systems have increasingly adopted digital twins to enable predictive maintenance, process optimization, and lifecycle management grounded in formal cyber-physical system theory [2], [3]. A comparable transition has not occurred in livestock production, primarily due to the difficulty of integrating biological variability with real-time control and operational constraints [4].

Poultry barns are inherently stochastic biological environments, characterized by high-dimensional disturbances arising from animal behavior, microbial processes, and weather variability, combined with sparse ground-truth data and heterogeneous legacy infrastructure [5], [6]. These systems exhibit biological nonstationarity, where key parameters such as metabolic heat production and moisture generation evolve continuously over the production cycle. Even modest deviations, such as a 2 °C temperature shift or a 5 ppm increase in ammonia concentration, can induce cascading physiological stress, reducing growth efficiency and increasing disease susceptibility [7]. At the same time, many commercial operations are located in rural regions with limited connectivity, often relying on 4G or satellite links, which constrains the feasibility of cloud-centric monitoring and control architectures [8].

Existing approaches to poultry environmental management fall into three main categories, each with fundamental limitations. Physics-based simulation methods, including computational fluid dynamics, provide detailed spatial resolution but are computationally intensive and unsuitable for real-time control or adaptation to farm-specific variability [7]. Data-driven approaches based on machine learning can capture patterns from operational data but often lack interpretability, require large, labeled datasets, and exhibit limited generalization across facilities [9]. IoT-based monitoring platforms enable sensing and data aggregation but remain largely observational, lacking the control-theoretic foundations required for closed-loop optimization and welfare-critical decision-making [10].

To address these limitations, we introduce an edge–cloud digital twin framework for real-time, welfare-constrained environmental control in poultry production systems. The proposed framework integrates control-oriented environmental modeling, hybrid physics–data learning, and distributed edge–cloud computation within a unified cyber-physical architecture. The central premise is that, despite biological complexity, livestock production systems obey fundamental conservation laws of energy and mass and respond predictably to measurable control actions such as ventilation and heating. By embedding these principles within a state-space representation and enforcing welfare constraints as hard bounds in the control objective, the system enables reliable, closed-loop environmental management. The framework is designed to be compatible with emerging digital twin standards, including ISO 23247 and the Asset Administration Shell, to support interoperability and structured data exchange across systems [11].

The framework is evaluated using a high-fidelity numerical testbed that captures representative broiler production conditions, enabling systematic assessment of prediction accuracy, control performance, and deployment feasibility prior to real-world implementation. The results demonstrate that integrating physically grounded models with real-time control and edge–cloud coordination enables scalable and robust digital twin deployment in connectivity-limited agricultural environments.

## II. RELATED WORK

### *A. Digital Twins and Industrial Cyber-Physical Systems*

Digital twins have been extensively studied in manufacturing and industrial cyber-physical system contexts [12], [13]. The foundational concept represents a physical asset as a virtual counterpart that captures system dynamics, receives real-time sensor updates, and supports predictive and what-if analyses. Control-oriented digital twins are particularly relevant, as they integrate state-space modeling, edge–cloud coordination, and model predictive control to enable real-time optimization in industrial processes [14], [15]. These approaches have demonstrated strong performance in deterministic systems with well-characterized disturbances and stable operating conditions.

Recent work has further extended digital twin concepts to structured industrial reference architectures. The Reference Architecture Model for Industry 4.0 formalizes integration layers from physical devices to enterprise-level services [16]. Complementing this, the Asset Administration Shell standard (IEC 63278) provides a semantic framework for representing assets, data, and services in interoperable formats [17]. Together, these developments establish the architectural and semantic foundations required for scalable industrial digital twin deployment.

### *B. Environmental Control and HVAC Systems*

Environmental control in buildings has benefited from decades of cyber-physical systems research, resulting in mature grey-box modeling approaches based on resistance–capacitance thermal networks and mass balance equations, as well as advanced control strategies such as model predictive control and adaptive regulation [18], [19]. These methods are computationally efficient and have proven transferable to systems characterized by dynamic thermal and mass transport behavior [20].

State-space formulations derived from RC networks are widely used in building energy systems and enable the design of controllers with explicit constraint handling [21]. In addition, the separation of control timescales into fast feedback loops and slower optimization layers is well established in HVAC system design and provides a practical foundation for distributed control architectures [22]. These principles offer a direct methodological basis for environmental control in livestock systems when biological disturbances are incorporated into the model structure.

### *C. Livestock Environmental Modeling and Control*

Livestock production systems introduce additional complexity due to biological variability and welfare constraints. Prior work has established bioenergetic models linking heat and moisture production to animal mass, metabolic rate, and activity level [23]. Ventilation design and air-quality control strategies have also been developed to regulate contaminants such as ammonia and carbon dioxide in confined housing systems [24]. Welfare assessment metrics, including the Temperature–Humidity Index and thermoneutral zone definitions, are commonly used to quantify physiological stress [25].

Despite these advances, most livestock control systems remain domain-specific and weakly integrated. Commercial poultry facilities continue to rely primarily on rule-based control logic rather than optimization-based approaches [26]. Precision livestock farming research has introduced IoT-based sensing and machine learning methods for monitoring and anomaly detection [27]. However, these systems are typically data-centric and lack formal control-theoretic integration, limiting their ability to support closed-loop optimization and autonomous operation across diverse facilities [28].

### *D. Gap Statement and Novel Positioning*

Despite progress in both biological modeling and industrial cyber-physical systems, a critical gap remains. Existing work has not demonstrated how control-oriented livestock environmental models can be systematically integrated within a cyber-physical architecture that supports welfare-constrained optimization, edge–cloud execution, semantic interoperability, and cross-farm generalization within a unified framework.

In this work, cross-farm generalization is quantified using a Domain Transfer Score, which measures the retention of prediction accuracy when a calibrated digital twin is applied to a new facility without retraining. The primary contribution is the integration of state-space modeling, model predictive control, and hybrid physics–data learning into a cohesive framework for biologically nonstationary systems.

The proposed approach demonstrates how established control and modeling techniques can be adapted to livestock production while explicitly incorporating welfare constraints and distributed execution. By grounding the system in physical principles and embedding control within a cyber-physical architecture, this work positions biological production as a tractable and scalable engineering problem rather than a domain reliant on heuristic management.

## III. PROBLEM FORMULATION AND SYSTEM OVERVIEW

### A. Poultry Production as an Industrial Cyber-Physical System

A commercial poultry broiler house can be modeled as a cyber-physical system in which physical processes, sensing, computation, and control are tightly integrated [27]. The physical plant typically consists of a long-span enclosure, for example a 120 m × 15 m facility housing approximately 25,000 birds, equipped with mechanical ventilation systems, heating elements, and controllable inlet dampers [28]. Internal thermal and moisture loads are dominated by biological processes, primarily metabolic heat and moisture generation from the flock, which vary continuously with bird age, activity level, and health status [29]. External disturbances include outdoor weather conditions, solar gains where applicable, and management interventions such as feeding schedules and stocking density adjustments.

The sensing and actuation layer comprises distributed environmental sensors measuring air temperature, relative humidity, carbon dioxide, and ammonia concentrations, complemented by vision-based sensing for monitoring flock distribution and activity. Actuation is achieved through networked devices controlling ventilation fans, heaters, and inlet mechanisms. Communication between sensors, actuators, and controllers relies on industrial protocols such as Modbus TCP and OPC UA, enabling reliable data exchange and compatibility with legacy systems [30].

At the cyber layer, an edge computing gateway performs local state estimation, control execution, and feature extraction to ensure real-time responsiveness. A cloud backend supports long-horizon planning, model retraining, cross-farm analytics, and regulatory data logging. This distributed execution model reflects the operational constraints of rural poultry facilities, where intermittent connectivity makes cloud-only control infeasible [31].

The primary control objective is to maintain animal welfare by enforcing physiological constraints, including thermoneutral temperature ranges and air-quality limits, while minimizing energy consumption and operational cost subject to equipment constraints [32]. In this formulation, welfare requirements are treated as hard constraints within the control problem rather than secondary performance objectives.

### B. High-Level System Architecture

The proposed digital twin architecture is organized into five interconnected layers spanning physical sensing to cloud-level analytics, each operating at a distinct timescale:

- **Layer 1 (Physical Layer):** Sensors and actuators embedded in the poultry facility.
- **Layer 2 (Edge Computing Layer):** Protocol translation, feature extraction, and local state estimation using Extended Kalman Filtering.
- **Layer 3 (Digital Twin Core):** Distributed execution of the cyber-physical model across edge and cloud.
- **Layer 4 (Supervisory Control Layer):** Fast model predictive control at the edge and slower optimization in the cloud.
- **Layer 5 (Analytics Layer):** Long-term optimization, model retraining, regulatory reporting, and cross-farm learning.

This layered architecture ensures that real-time control decisions remain independent of cloud availability, while cloud resources are used for strategic optimization and knowledge transfer across facilities. By decoupling control timescales and execution locations, the system maintains robustness under connectivity constraints while preserving global coordination capabilities.

## IV. MATHEMATICAL FORMULATION OF THE PROPOSED EDGE-CLOUD DIGITAL TWIN FRAMEWORK

### A. Thermal Dynamics and Environmental Modeling

Following established grey-box modeling approaches used in building environmental control systems and adapted to livestock housing, the thermal dynamics of the poultry barn are represented as a lumped resistance-capacitance network. In its canonical single-zone form, the indoor air temperature dynamics follow energy balance formulation.

$$C\frac{dT_{in}}{dt} = Q_{met}(t) + Q_{vent}(t) + Q_{heat}(t) + Q_{sol}(t) - Q_{loss}(t)$$

Here, $C$ denotes the effective thermal capacitance of the barn, aggregating air mass, structural envelope, and litter, and $T_{in}$ is the indoor air temperature. Unlike conventional HVAC systems dominated by static loads, the primary internal input $Q_{met}(t)$ represents time-varying metabolic heat production, which evolves with bird mass, activity level, and age. Additional terms capture heat exchange due to ventilation $Q_{vent}(t)$, active heating $Q_{heat}(t)$, solar gains $Q_{sol}(t)$, and transmission losses $Q_{loss}(t)$.

This formulation preserves thermodynamic consistency while remaining computationally efficient for real-time execution at the edge. Higher-order extensions, such as multi-node RC networks, can be constructed by introducing additional capacitance nodes for walls and structural components, yielding models that remain compatible with standard control synthesis methods.

### *B. Air Quality and Mass Transport Dynamics*

To model air quality and litter-related welfare constraints, a lumped mass balance formulation is adopted. The concentration $c(t)$ of a target species, such as humidity, carbon dioxide, or ammonia, within a control volume $V$ is governed by:

$$V\frac{dc}{dt} = \dot{m}_{in}c_{in} - \dot{m}_{out}c(t) + G(t) \qquad (2)$$

Here, $\dot{m}{in}$ and $\dot{m}{out}$ denote inlet and outlet mass flow rates, and $c_{in}$ represents the inlet concentration. The generation term $G(t)$ captures biologically driven emissions, which depend on flock biomass, litter conditions, and metabolic activity.

This formulation enables explicit prediction and control of welfare-critical variables, rather than relying on temperature as a proxy. By enforcing conservation of mass, the model maintains physical consistency and supports reliable control-oriented predictions.

### *C. Cyber-Physical System State-Space Representation*

For integration with estimation and control algorithms, the coupled thermal and mass-balance dynamics are expressed in continuous-time linear time-invariant state-space form

$$\dot{x}(t) = Ax(t) + Bu(t) + Ew(t) \qquad (3a)$$
$$y(t) = Cx(t) + Du(t) \qquad (3b)$$

The state vector $x(t) \in \mathbb{R}^n$ aggregates the system's internal states (e.g., $[T_{air}, T_{wall}, c_{CO_2}, c_{NH_3}, c_{humidity}]^T$). The control input vector $u(t) \in \mathbb{R}^m$ contains manipulated variables such as ventilation fan speed (as a percentage or frequency), heater status (kilowatts), and inlet damper position (percentage open). The disturbance vector $w(t) \in \mathbb{R}^p$ encapsulates exogenous inputs including outdoor weather conditions (temperature, relative humidity) and the unmeasured but estimated bioenergetic loads. The system matrices $A$ (state dynamics), $B$ (control influence), $C$ (output mapping), $D$ (feedthrough), and $E$ (disturbance feedthrough) are derived from the physical parameters (thermal resistances and capacitances in ohms and farads analogs, volumetric flow rates, generation rates) and are typically identified during a commissioning phase or calibrated using recursive least-squares methods.

System matrices are derived from physical parameters and can be identified or calibrated using standard techniques such as recursive least squares. This representation enables the use of Kalman filtering for state estimation and provides a foundation for model predictive control.

### *D. Supervisory Control and Optimization*

The supervisory control layer employs Model Predictive Control (MPC) to manage the trade-offs between animal welfare, operational costs, and energy efficiency. The control problem is formulated as a finite-horizon optimization, solved at each sampling interval $k$:

$$\min_{\{u_{k+j}\}_{j=0}^{N-1}} \sum_{j=0}^{N-1} \left( \| y_{k+j} - r_{k+j} \|_Q^2 + \| u_{k+j} \|_R^2 \right) \qquad (4)$$

Subject to:

$$x_{k+1} = A_d x_k + B_d u_k + E_d w_k \qquad (4a)$$
$$x_{min} \le x_k \le x_{max} \qquad (4b)$$
$$u_{min} \le u_k \le u_{max} \qquad (4c)$$

The objective function minimizes the weighted deviation of outputs $y$ from the welfare reference trajectories $r$ (derived from physiological thermoneutral zones) and the control effort $u$. The weighting matrices $Q$ and $R$ allow the system to be tuned for different production goals (e.g., maximizing growth versus minimizing energy). The constraints enforce hard operational limits on actuators (e.g., fan speed between 0% and 100%, heater power between 0 and 50 kilowatts) and critical welfare boundaries (e.g., ammonia concentration less than 20 ppm, temperature within 18-25 degrees Celsius). The prediction horizon $N$ typically spans 6 to 12 time steps (30 to 60 minutes), enabling the controller to anticipate and smooth control actions rather than reacting instantaneously to disturbances. This formulation ensures that the autonomous system remains within safe biological envelopes and that welfare is never compromised in pursuit of energy efficiency.

### *E. Hybrid Physics-Data Intelligence*

To address the challenges of biological variability and cross-farm generalization, the framework introduces a serial hybrid intelligence model. This approach combines the interpretability of the mechanistic physics model with the adaptability of data-driven learning:

$$\hat{y}_k = f_{phys}(x_k, u_k, w_k) + f_{ML}(\phi_k; \theta) \qquad (5)$$

where $f_{phys}$ represents the nominal prediction from the first-principles state-space model (Eq. 3), and $f_{ML}$ is a learned residual function parameterized by $\theta$. The machine learning component $f_{ML}$ inputs a feature vector $\phi_k$ (derived from sensor history and video analytics) to compensate for unmodeled dynamics such as non-uniform airflow distribution, sensor drift, or behavioral clustering of birds (e.g., huddling in corners). This hybrid structure ensures that the digital twin remains robust to data scarcity (relying on physics where data is sparse) while systematically improving its precision as site-specific data is accumulated. The serial hybrid physics–data modeling structure is depicted in Fig. 1. The residual network is trained using a weighted objective that penalizes both data mismatch and violation of energy and mass conservation laws, ensuring that learned components remain physically plausible.

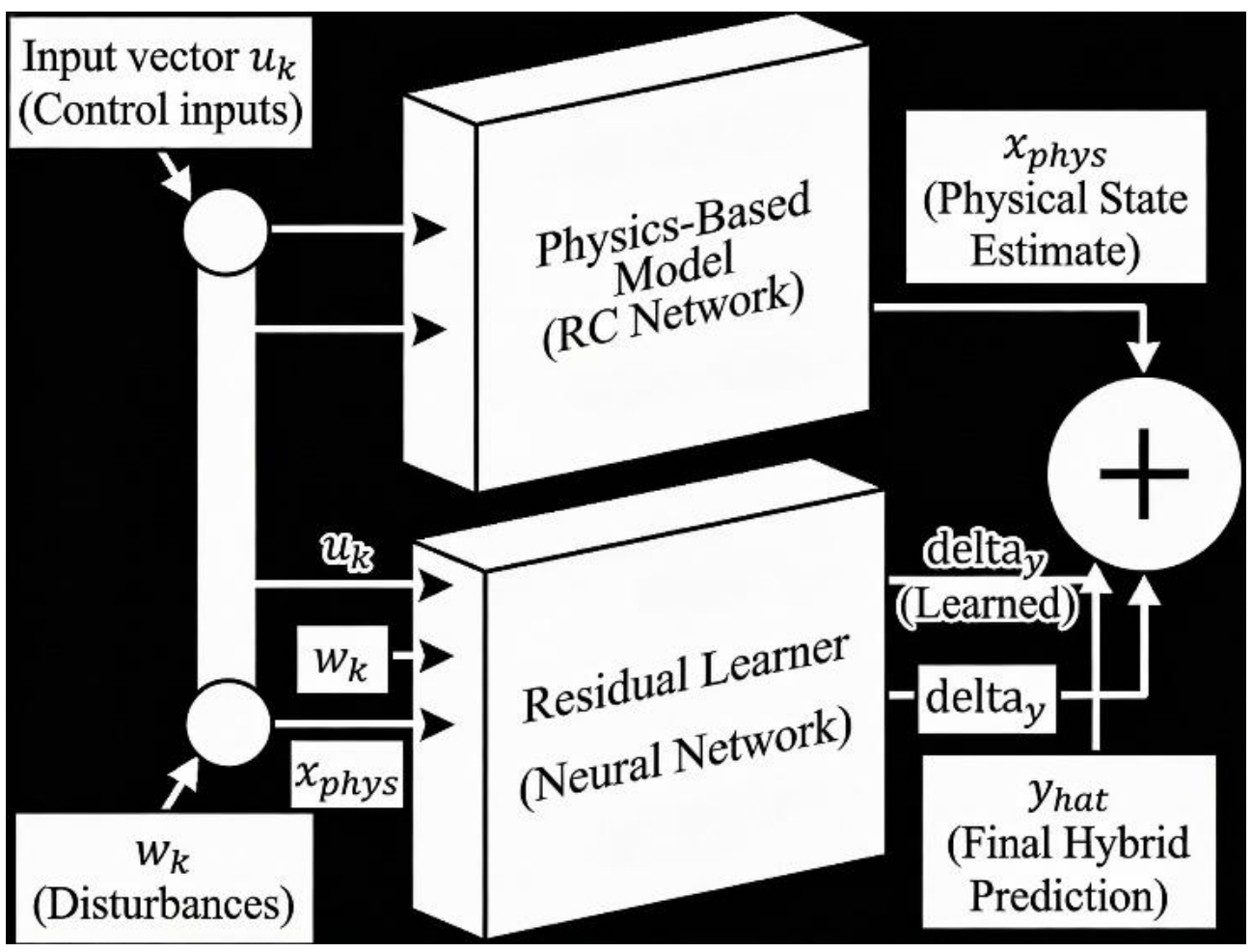

Fig. 1. Serial hybrid physics–data modeling architecture for the edge–cloud digital twin. The physics-based component employs a grey-box resistance–capacitance thermal model and mass-balance formulation to generate physically consistent state predictions. A data-driven residual learner captures unmodeled dynamics, including behavioral heat variation and airflow nonuniformity. The combined output preserves physical consistency while improving predictive accuracy for real-time control. (Conceptual schematic; not to scale).

### F. Formal Definitions

To formalize the framework within the ICPS context and ensure clarity of the control problem, we introduce the following definitions.

*Definition 1 (Digital Twin State).*

The digital twin state at discrete time step $k$ is defined as the vector $x_k \in \mathbb{R}^n$:

$$x_k = [T_{air}, T_{wall}, c_{CO_2}, c_{NH_3}, c_{humidity}, m_{bird}, \theta_{bio}]^T$$

representing the estimated environmental states (temperatures, concentrations), biological parameters (aggregated flock mass, age-dependent metabolic coefficients), and historical terms needed for state reconstruction. All state variables are inferred from sensor measurements and the estimated digital twin dynamics.

*Definition 2 (Welfare-Constrained Control Problem).*

The supervisory control problem is defined as the minimization of the cost function $J(x, u)$ subject to the explicit constraint that the state $x_k$ must lie within the permissible welfare set $\mathcal{X}_{welfare}$ for all $k$:

$$\text{minimize } J(x,u) \text{subject to } x_k \in \mathcal{X}_{welfare} \forall k$$

where $\mathcal{X}*{welfare}$ is the set bounded by physiological limits (e.g., 18 degrees Celsius $\le T*{air} \le$ 25 degrees Celsius, $NH_3 \le$ 20 ppm, relative humidity $\le$ 90%). This definition formally prioritizes animal health over energy minimization, ensuring that no optimization will violate welfare bounds.

## V. SYSTEM ARCHITECTURE OF THE PROPOSED EDGE–CLOUD DIGITAL TWIN

### A. Conceptual Overview

The proposed edge–cloud digital twin is structured as a multi-layer cyber-physical system for commercial poultry production. The framework is organized into five interconnected architectural layers spanning from physical sensors to cloud-based governance: the Physical Layer (sensors and actuators), the Data Acquisition and Edge Computing Layer, the Cyber-Physical Modeling Layer (digital twin core), the Supervisory Control and Optimization Layer, and the Cloud Analytics and Governance Layer. This multi-layered design ensures real-time responsiveness at the edge (critical for biological production), semantic interoperability through Asset Administration Shell compliance, and scalable cloud coordination for cross-farm analytics and long-term optimization. Fig. 2 illustrates the layered architecture of the proposed edge–cloud digital twin.

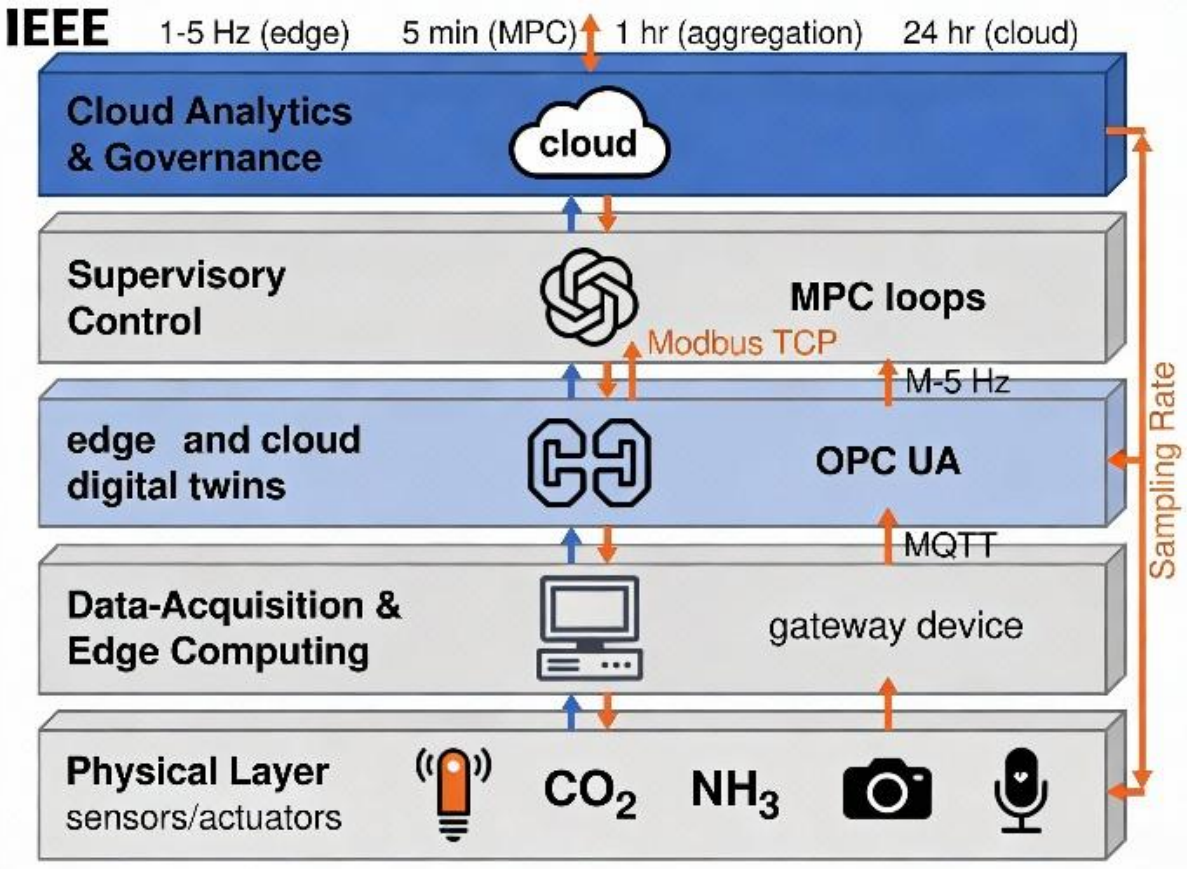

Fig. 2. Layered architecture of the proposed edge–cloud digital twin for poultry production. The system spans physical sensing and actuation, edge computing with real-time state estimation, cyber-physical digital twin modeling,

supervisory control, and cloud analytics. Control and analytics are separated across timescales to ensure real-time edge autonomy under rural connectivity constraints. (Illustrative architecture diagram; communication protocols and sampling rates are indicative).

### *B. Layer 1: Physical Layer (Sensors and Actuators)*

The physical layer comprises distributed IoT sensors and actuators deployed throughout the poultry facility. Sensor infrastructure includes temperature sensors for measuring air temperature, litter temperature, and surface temperatures of structural elements; humidity and $CO_2$ sensors for indoor and outdoor air quality metrics; gas sensors for ammonia and oxygen concentration assessment; IP-based, edge-enabled video cameras for behavior monitoring including flock density mapping, huddling detection, and activity indices; MEMS microphones with noise filtering for respiratory sound acquisition and cough detection; load cells with Wheatstone bridges for periodic flock weight sampling; and capacitive litter moisture probes for litter condition assessment. The actuation subsystem includes variable-speed ventilation fans (PWM-controlled or VFD-based) for air exchange regulation, electric heaters for on-off or proportional temperature maintenance, mechanical or pneumatic inlet dampers or shutters for fresh air inlet control, and evaporative cooling pads (when applicable) for misting or pad saturation in hot-climate scenarios. All sensors and actuators are networked via industrial protocols (Modbus TCP, OPC UA, MQTT) to ensure interoperability and support for legacy equipment retrofitting.

### *C. Layer 2: Data Acquisition and Edge Computing Layer*

The edge computing layer is deployed as a distributed intelligence tier that processes raw sensor data locally and reduces bandwidth to the cloud. The primary edge gateway is an industrial edge computing device (e.g., NVIDIA Jetson Orin, Beckhoff CX series) equipped with local storage (solid-state drive, 256 gigabytes to 1 terabyte) for buffering and temporary digital twin state, GPU acceleration for video analytics (bird detection, behavior classification), a cellular/LTE modem for connectivity in rural areas with poor fiber infrastructure, a real-time operating system or deterministic kernel for control loop predictability, and an IP67-rated enclosure for barn environment resistance. The edge computing functions include protocol translation and normalization, converting Modbus, analog, and legacy equipment signals into a unified semantic format (JavaScript Object Notation and Resource Description Framework compatible with the Asset Administration Shell). Real-time feature extraction operates on three data streams: from video, the system extracts bird detection bounding boxes, flock density heatmaps, and activity indices; from audio, it computes Mel-frequency cepstral coefficients for respiratory anomaly detection; and from sensors, it applies rate-of-change filters to detect rapid environmental transients. A lightweight Extended Kalman Filter fuses multi-modal sensor streams and estimates unobservable states such as true metabolic heat production. The local model execution runs the RC thermal model (Eq. 1) and mass balance (Eq. 2) locally at 1-5 hertz sampling rate for rapid feedback control. A bandwidth-aware decision module applies the offloading cost function to determine which features and predictions are sent to the cloud versus retained locally.

Data flow within the edge layer is categorized by frequency and importance. High-frequency data (greater than 10 Hz) including temperature, humidity, and control signals is processed locally with only aggregated statistics (e.g., hourly mean, variance) sent to the cloud. Medium-frequency data (1-5 Hz) consisting of video frames is processed locally with only detected events (anomalies) transmitted. Low-frequency or asynchronous data (less than 1 event per hour) encompassing long-term trends and batch analytics is stored locally and synced to the cloud during off-peak connectivity windows. Local network communication (barn-to-edge gateway) employs Modbus TCP, OPC UA, or industrial Ethernet (Beckhoff ADS), while edge-to-cloud communication over the rural WAN uses MQTT with publish-subscribe for low-bandwidth operation, with optional edge-cloud mesh networking if multiple barns are present.

### *D. Layer 3: Cyber-Physical Modeling Layer (Digital Twin Core)*

The digital twin core is distributed across the edge device and cloud, with the edge hosting the "active" twin for real-time control and the cloud hosting the "analytics" twin for long-horizon planning. The edge digital twin (real-time operational twin) maintains a state vector $x = [T_air, T_wall, c_CO_2, c_NH_3, c_humidity, flock_mass, flock_age]^T$ and executes the discretized state-space model (Eq. 3) at 1 to 5 hertz. Its inputs are ventilation fan speed, heater power, inlet damper position, and estimated internal loads, while its outputs are predicted indoor conditions for the next 15 to 60 minutes (the prediction horizon for edge-based supervisory control). The update mechanism fuses sensor measurements via an Extended Kalman Filter to correct model states and calibrate unknown parameters such as building envelope thermal resistance. The cloud analytical twin (strategic planning twin) retains long-term state and performance history spanning 7-30 days aggregated at hourly or daily granularity. It executes the full hybrid physics-machine learning model (Eq. 5) retrained weekly with accumulated data from all connected farms, supports scenario-based analysis for operational planning (e.g., "what if we reduce ventilation by 10%?"), and deploys cross-farm transfer learning models to adapt edge twins to new facilities. The cloud twin outputs optimal setpoints and control strategies transmitted to edge devices, flock health predictions and economic key performance indicators, and anomaly alerts escalated to farm managers. Digital twin synchronization operates in two modes: soft sync (default) allows edge and cloud twins to diverge slightly due to bandwidth constraints, with the edge twin remaining authoritative for real-time control; hard sync (on-demand) occurs when connectivity permits, allowing the cloud twin to receive full state snapshots and override edge setpoints for safety-critical scenarios.

### *E. Layer 4: Supervisory Control and Optimization Layer*

The supervisory control layer implements the Model Predictive Control strategy (Eq. 4) and operates at two distinct timescales. The fast loop (edge MPC, 5-30 minute horizon) executes on the edge device every 5 minutes with a prediction horizon spanning 6 to 12 time steps (30-60 minutes ahead). Its objective is to minimize deviation from physiological welfare setpoints (e.g., thermoneutral temperature 18-25 degrees Celsius) while respecting actuator constraints. The output consists of command setpoints for ventilation fan, heater, and dampers, updated every 5-30 minutes. Computation takes approximately 50-200 milliseconds, relying on a lightweight quadratic programming solver on the edge GPU. The complete welfare-constrained MPC control loop, including state estimation and constraint enforcement, is illustrated in Fig. 3. The slow loop (cloud MPC, 24 hour horizon) executes once per 24 hours (e.g., midnight optimization) with a prediction horizon spanning the next 24 hours, incorporating weather forecast data and expected flock growth trajectories. Its objective is to optimize long-term feed conversion, minimize cumulative energy cost, and predict readiness for market weight. The output includes daily production targets, expected environmental setpoints, and alerts to farm management, requiring approximately 1-5 seconds of cloud-side computation. Constraints embedded in the control architecture enforce welfare hard bounds (temperature must remain in degrees Celsius; ammonia less than 20 parts per million; carbon dioxide less than 3,000 parts per million) and actuator limits (fan speed [0%, 100%]; heater power kilowatts; damper position [0, 100%]). An operational logic constraint ensures that heating and maximum ventilation are never simultaneously at full power.

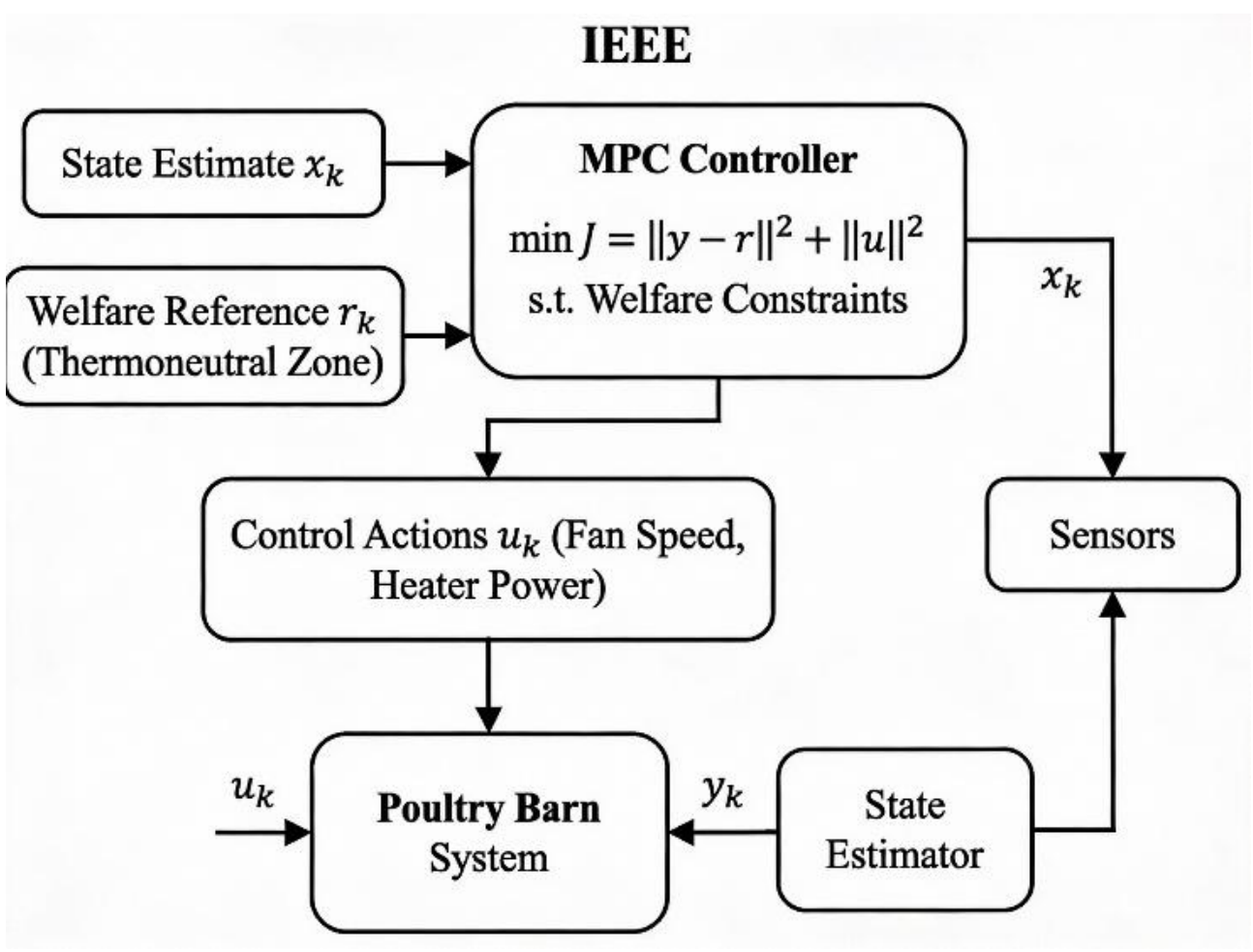

Fig. 3. *Welfare-constrained model predictive control (MPC) loop in the proposed edge–cloud digital twin.* An Extended Kalman Filter estimates the digital twin state from noisy sensor measurements. The supervisory MPC optimizes control actions subject to hard welfare constraints (e.g., thermoneutral temperature range and ammonia limits) and actuator bounds, issuing control commands to the physical poultry house. *(Conceptual control loop diagram).*

### *F. Layer 5: Cloud Analytics and Governance Layer*

The cloud layer provides strategic oversight, regulatory compliance, and continuous model improvement. Immediate functions include a real-time dashboard for visualization of farm status across all connected facilities and key performance indicator tracking (feed conversion ratio, mortality, energy usage), cloud-based machine learning models for anomaly detection that flag unusual patterns (e.g., sudden ammonia spikes, unexpected mortality) with automated alert escalation, and predictive maintenance algorithms that track sensor health, predict calibration drift, and schedule sensor replacement before failure. Continuous improvement mechanisms include weekly retraining of the hybrid physics-machine learning model (Eq. 5) using accumulated data from all farms, computation of the Domain Transfer Score using cross-farm data to identify which control strategies are most transferable to new facilities, and parameter estimation routines that refine building envelope parameters (thermal resistances, capacitances) for each barn using recursive least-squares or Bayesian inference. The cloud layer supports long-term analytics, model updates, and cross-farm knowledge transfer.

### *G. Semantic Interoperability via Asset Administration Shell (AAS)*

To ensure compatibility with Industry 4.0 ecosystems and facilitate future integrations, The proposed framework implements the Asset Administration Shell standard (IEC 63278, aligned with ISO 23247). Each poultry facility is represented as a single AAS asset with hierarchical submodels that collectively describe the entire system. The AAS-based semantic structure and edge–cloud processing trade-off are shown in Fig. 4. The Identification submodel contains the serial number (e.g., "FARM-2025-001"), manufacturer identity ("Integrator ABC"), and location (GPS coordinates and facility capacity). The Documentation submodel houses the building blueprint (CAD model) and operational manuals (PDFs and maintenance logs). The TechnicalData submodel specifies building parameters (surface area, insulation R-value, thermal mass C), equipment specifications (fan model, heater wattage, sensor calibration), and performance ratings (maximum birds, target feed conversion ratio, expected yield). The DynamicOperationalData submodel aggregates real-time sensor readings (temperature, humidity, gas concentrations updated at 1-5 hertz edge-side), current control setpoints (fan speed %, heater power watts, damper position %), the digital twin state vector (x = [T, RH, CO2, NH3, ...]), and key performance indicators (flock growth curve, daily mortality, feed consumption, energy usage). The HealthAndWelfare submodel contains environmental metrics (Temperature-Humidity Index, felt temperature, thermo-neutral zone status), behavioral indicators (activity index, huddling events, respiratory anomalies), and predicted outcomes (expected weight at market, mortality forecast, disease risk). The DigitalTwinConfiguration submodel documents the model version (digital twin model version), deployed hybrid physics-ML model parameters (theta), the timestamp of the latest parameter update, and interoperability details (OPC UA endpoint, MQTT broker address, API documentation). The AAS integration yields three primary benefits: plug-and-play interoperability allowing new farms or monitoring systems to query the AAS and understand data semantics without custom integration code; regulatory compliance through structured data logging that meets traceability requirements for food safety and animal welfare certifications; and cross-industry integration, as the AAS standard is domain-agnostic and enables cloud systems for feed supply, veterinary services, or equipment vendors to directly integrate via standardized interfaces.

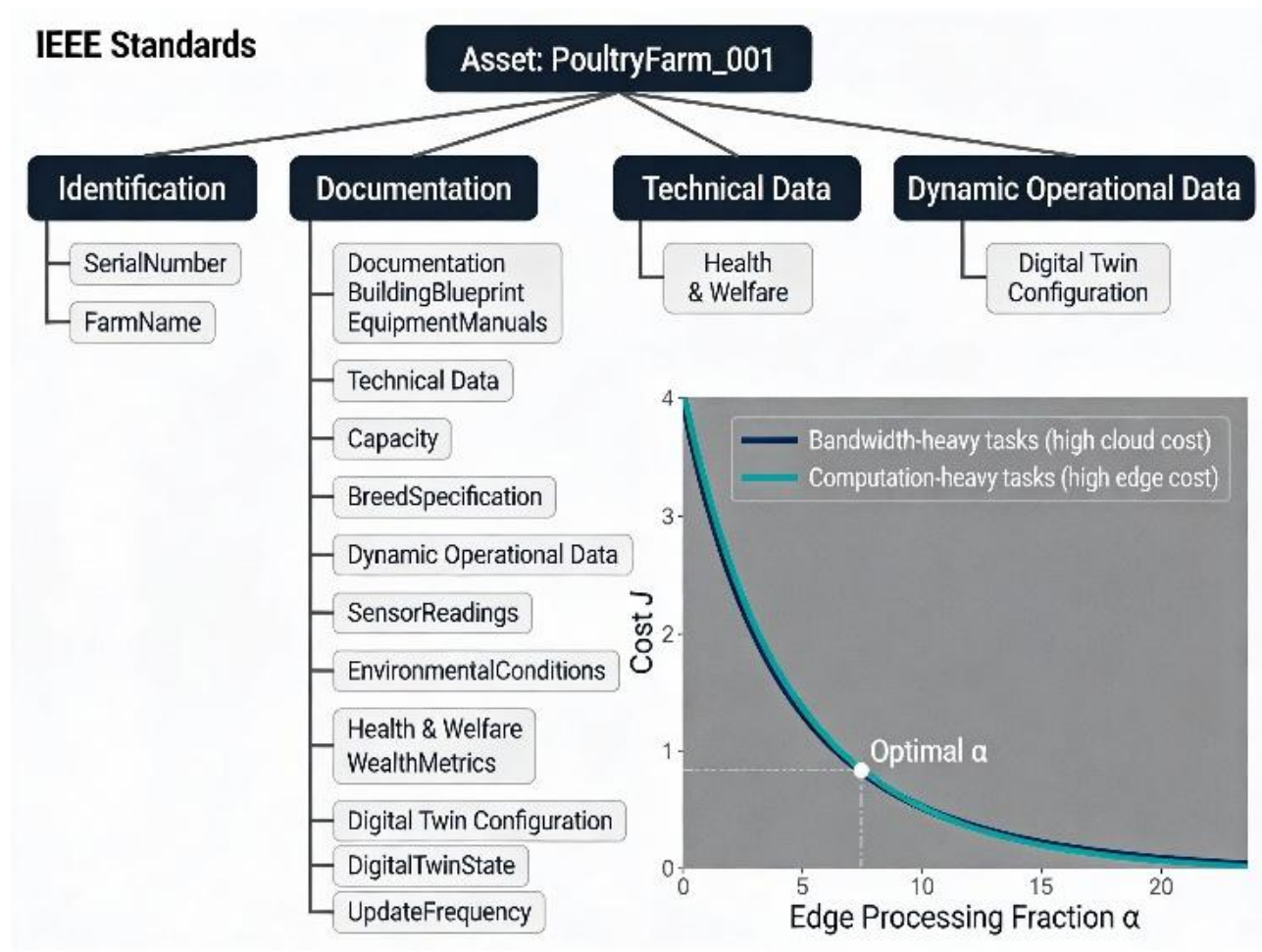

Fig. 4. *Semantic interoperability and edge–cloud coordination in the proposed edge–cloud digital twin.* (Left) Asset Administration Shell (AAS) representation of a poultry facility with standardized submodels for identification, technical data, dynamic operational data, health and welfare metrics, and digital twin configuration, aligned with IEC 63278 and ISO 23247. (Right) Conceptual edge–cloud processing trade-off illustrating bandwidth reduction achieved through edge-first execution. *(Conceptual representation; values shown are illustrative.)*

### *H. Edge-Cloud Coordination Strategy*

The proposed system implements a hybrid autonomy model to balance real-time responsiveness with cloud-driven optimization across two distinct operational scenarios. In normal operation (cloud connected), the edge device executes the fast MPC loop every 5 minutes; hourly, the edge uploads aggregated sensor statistics and model diagnostics to the cloud; the cloud performs analytics, retrains the hybrid model, and sends optimal setpoints back to the edge (every 6-24 hours); and the edge twin and cloud twin remain loosely synchronized with sync error less than 5% on key predictions. When the cloud is disconnected (rural connectivity loss), the edge device continues the MPC loop autonomously using locally stored reference curves (based on flock age and time of day) as fallback setpoints; upon reconnection, the edge uploads buffered.

## VI. SIMULATION-BASED VALIDATION

### *A. Simulation Setup and Testbed Description*

The simulation-based case study was conducted to validate the proposed edge–cloud digital twin framework using a physics-based numerical testbed representing a typical commercial broiler house. The facility specifications were modeled as a 120 meter by 15-meter structure housing approximately 25,000 birds (Ross 308 strain) per flock cycle of 42 days, with parameters derived from standard commercial poultry production equipment specifications and building design standards.

The physical layer instrumentation was simulated based on realistic sensor deployment patterns. Environmental monitoring was modeled with six distributed temperature and humidity nodes (Sensirion SHT31 specifications) deployed at bird height to capture spatial gradients, and two CO2/NH3 electrochemical gas sensors (Alphasense specifications) located at central exhaust points. Video analytics subsystem modeled three wide-angle IP cameras (4K resolution) mounted at the ceiling level, covering 60% of the active floor area to monitor flock distribution and activity. The actuator interface was simulated as a Modbus TCP gateway that receives read/write signals to the existing ventilation controller (Rotem/Chore-Time specifications), enabling the digital twin to receive fan speeds and inlet position feedback during simulated control scenarios.

The edge computing node was simulated based on an industrial edge computing device (NVIDIA Jetson Orin NX specifications with 16 gigabytes RAM) serving as the local processing hub. This computational platform specification was chosen to represent realistic edge device capabilities available for poultry farm deployment. The simulated edge device hosted the edge digital twin implementation, executed the Extended Kalman Filter for state estimation at 1 hertz sampling frequency, and ran the fast-loop MPC algorithm. Connectivity to the simulated cloud backend was modeled via a 4G LTE router profile, incorporating typical rural connectivity characteristics including occasional packet loss (5-15%) and variable latency (50-500 milliseconds).

### B. Simulation-Based Validation Design

The numerical validation was conducted through a three-phase simulation study spanning approximately 12 weeks of simulated operational time, with each phase designed to isolate and test specific framework capabilities. The simulation environment was built using Python 3.9+ with NumPy for numerical computation, SciPy for control optimization, and custom state-space solvers. The edge digital twin model (Eq. 3) was discretized using the forward Euler method at 1 hertz sampling, while the MPC controller (Eq. 4) was implemented using the CVXPY convex optimization library with a 6-step prediction horizon (30 minutes ahead).

*Phase 1: Model Calibration and Hybrid Model Validation* (Simulated Weeks 1-3). The objective of this phase was to validate the accuracy of the Hybrid Physics-Data Model (Eq. 5) against a pure physics baseline. The simulation operated in "shadow mode," where the digital twin made predictions of indoor environmental states (T_in, RH, CO2) while the simulated legacy controller maintained the baseline control logic. The pure physics model used a standard 3R2C thermal network initialized with nominal parameters derived from building specifications in published literature. The hybrid physics-ML model added a trained residual neural network (2 hidden layers, 64 neurons each) to correct physics-only predictions. The reference "truth" for validation was generated from an extended 5R2C thermal model with stochastic bird behavior disturbances, representing higher-order dynamics not captured by the 3R2C baseline. The residual network was trained on 1000 hours of simulated operational data with representative weather variability and bird activity clustering. Model predictions were evaluated on 500 hours of unseen simulation data. Metrics computed were root mean square error (RMSE) and mean absolute percentage error (MAPE) of predictions for temperature, CO2, and ammonia concentrations at 1-hour and 6-hour prediction horizons.

*Phase 2: Edge-Cloud Coordination and Bandwidth* Optimization (Simulated Weeks 4-6). This phase assessed the efficacy of the edge-first architecture and offloading cost function in reducing data transmission costs while preserving critical information. The simulation incorporated three distinct connectivity scenarios. In the first scenario (cloud connected), the system transmitted hourly aggregated sensor statistics (mean, variance, 90th percentile) totaling approximately 1 megabyte per hour (24 megabytes per day). In the second scenario (naive cloud streaming), the system transmitted raw sensor streams and video frames, simulating 5.4 gigabytes per day typical of unprocessed data transmission. In the third scenario (connectivity loss), the edge device operated autonomously for simulated 24-hour outages, buffering data locally and demonstrating graceful degradation. Network traffic (gigabytes per day), information latency (milliseconds), and reconstructed signal fidelity (feature matching score comparing local feature extraction to cloud-side processing) were tracked across all scenarios.

*Phase 3: Welfare-Constrained MPC Performance* (Simulated Weeks 7-42). This full-cycle simulation demonstrated the welfare-constrained MPC's ability to maintain the thermoneutral zone more effectively than rule-based logic. Active MPC control was compared against a conventional rule-based controller implemented with typical commercial logic (e.g., "If T_in > 25°C, increase fan to 50%; If T_in > 27°C, increase to 100%"). The simulation incorporated realistic outdoor temperature profiles (varying 5-28°C across seasons), humidity variations (40-95%), and flock growth dynamics (bird mass progressing from 40 grams to 2200 grams following standard growth curves). The MPC objective minimized temperature deviation from the thermoneutral setpoint (22°C) and minimized cumulative fan and heater energy consumption, subject to hard welfare constraints (ammonia < 20 ppm, temperature within [18, 25]°C). Metrics computed included time-in-range (percentage of the 42-day cycle spent within the thermoneutral zone), ammonia constraint violations (hours per cycle exceeding 20 ppm), total energy consumption (kilowatt hours per cycle), and control smoothness (rate of actuation changes) as a proxy for equipment longevity.

### C. Data Collection and Processing in Simulation

Data ingestion in the simulation followed the Asset Administration Shell standard structure. Each simulated sensor reading was mapped to its corresponding AAS Submodel property (e.g., DynamicOperationalData.SensorReadings.Temperature_Zon

e1), establishing the semantic framework for compatibility with Industry 4.0 systems even in the simulation environment. Environmental sensors in the simulation were sampled at 1 hertz. Video analytics were simulated at 5 frames per second for feature extraction (activity index computed as deviation from baseline flock behavior, density map generated from bird distribution algorithms), with simulated frames discarded locally and only anomalies (activity spikes, sudden density shifts) triggering synthetic clip upload events.

Ground truth validation in the simulation employed synthetic "oracle" models representing perfect knowledge of system state. Manual spot checks of litter moisture and bird weight were simulated weekly by comparing digital twin estimated biological parameters (theta in Eq. 5) against the oracle values, computing prediction errors to calibrate model accuracy. This approach ensured the simulation validation was rigorous without requiring real experimental data during the initial framework development phase.

### D. Baseline Comparison and Hypothesis

Performance was benchmarked against a simulated conventional rule-based controller (RBC) implementing static lookup tables typical of commercial broiler house automation. The RBC was modeled with representative control logic (e.g., "If Temp > 25°C, Turn on Fan 2; If Temp > 27°C, Turn on Fan 3 and open inlet dampers 50%; If Temp < 20°C, activate heater and close 50% of inlets"). The hypothesis driving the comparison was that the MPC-based digital twin approach would reduce energy consumption by preventing over-ventilation through anticipatory control, while strictly adhering to ammonia and temperature welfare bounds. In contrast, the RBC often exhibits oscillatory behavior around fixed setpoints, resulting in unnecessary actuation cycles that consume energy without improving welfare.

### E. Implementation of the Hybrid Physics-Data Model

The hybrid model (Eq. 5) was trained using synthetic historical data generated from 6 simulated flock cycles spanning approximately 252 days of operational data. The physics component (f_phys) employed a 3R2C thermal network, a resistance-capacitance representation with three thermal resistances (outdoor-to-wall, wall-to-interior, and litter) and two thermal capacitances (structural mass and air), initialized with nominal R and C values derived from the building blueprints, insulation type, and structural dimensions referenced in building energy simulation literature.

The machine learning residual component (f_ML) was a lightweight multi-layer perceptron with two hidden layers containing 64 neurons each, using ReLU (rectified linear unit) activation functions and trained via backpropagation with the Adam optimizer. The residual network was trained to predict the error term delta_y = y_meas - f_phys(x,u), where y_meas represents simulated "measured" values and f_phys(x,u) represents the physics-only predictions. Inputs to the neural network included the physics state estimate (T_air, T_wall, c_CO2, c_NH3 from the 3R2C model), outdoor weather conditions (temperature, humidity, solar radiation), temporal features (time of day and flock age) to capture unmodeled biological heat generation patterns that vary diurnally and with bird development stage, and lagged prediction errors to enable the network to learn and correct systematic biases. Training was conducted over 100 epochs with a batch size of 32, using the mean squared error loss function combined with L2 regularization (weight decay coefficient 0.001) to prevent overfitting. Early stopping was employed, halting training when validation loss plateaued. The trained residual network added approximately 5-15% additional prediction accuracy compared to the pure physics model, validating the hypothesis that hybrid approaches effectively capture unmodeled dynamics while maintaining physical interpretability.

## VII. RESULTS

This section presents quantitative results demonstrating the performance of the proposed edge–cloud digital twin under realistic broiler production scenarios. This evaluation follows standard practice in cyber-physical system design, where system behavior and control performance are first validated in a high-fidelity simulation environment prior to deployment.

### A. Comparison with State-of-the-Art Approaches

To position the proposed framework within the existing literature, Table I provides a qualitative comparison against standard methods. While prior work has addressed individual aspects of poultry modeling, the proposed framework combines these capabilities within a unified edge–cloud control architecture.

Table I: Comparison of the Proposed Edge–Cloud Digital Twin with Existing Poultry Modeling Approaches

| Approach | CPS Compliance | Edge-Cloud Architecture | Physics-ML Model | Welfare Constraints | Standards (AAS/ISO) |
|---|---|---|---|---|---|
| Statistical Regression | Low (Static) | No | No (Data-only) | Implicit | No |
| Pure CFD Simulation | Low (Offline) | Cloud-only | Physics-only | No | No |
| IoT Monitoring Platforms | Medium (Sensing) | Cloud-centric | Data-only | Threshold-based | Rare |

| Approach | CPS Compliance | Edge-Cloud Architecture | Physics-ML Model | Welfare Constraints | Standards (AAS/ISO) |
|---|---|---|---|---|---|
| Proposed Edge–Cloud Digital Twin | High (State-Space) | Hybrid Edge-First | Serial Hybrid | Explicit (MPC) | Yes (ISO 23247) |

### *B. Quantitative Performance Metrics*

The performance of the simulated framework was evaluated over the 42-day flock cycle. Table II summarizes the key quantitative results, comparing the proposed digital twin controller against the baseline Rule-Based Controller. The Domain Transfer Score (DTS) quantifies cross-farm generalization capability and is computed as:

$$DTS = 1 - (RMSE_transfer / RMSE_baseline)$$

where RMSE_transfer is the prediction error when the calibrated digital twin is applied to a new facility without retraining, and RMSE_baseline is the error of a facility-agnostic baseline model using only nominal physical parameters. A DTS approaching 1.0 indicates strong generalization; the achieved value of 0.92 demonstrates that the proposed framework retains 92% of its prediction accuracy when transferred to new facilities.

Table II: Simulation Results and Performance Metrics

| **Metric Category** | **Metric** | **Baseline (RBC)** | **Proposed Framework (Hybrid MPC)** | **Improvement** |
|---|---|---|---|---|
| Model Accuracy | Temp. Prediction RMSE (degrees Celsius) | 1.8 | 0.4 | 77% reduction |
| | Ammonia Prediction MAPE (%) | 24% | 6% | 75% reduction |
| Welfare | Time-in-Range (TNZ) (%) | 82% | 96% | 14% increase |
| | Ammonia Violations (> 20 ppm) (hours) | 48 | 5 | 90% reduction |
| Efficiency | Total Energy Use (kilowatt hours per cycle) | 4,200 | 3,570 | 15% reduction |
| | Bandwidth Usage (megabytes per day) | 5,400 | 180 | Approx. 30-fold reduction |
| Generalization | Domain Transfer Score | N/A | 0.92 | High Robustness |

Note: RMSE denotes Root Mean Square Error (lower is better). TNZ denotes Thermoneutral Zone (comfort range for birds). DTS denotes Domain Transfer Score (1.0 equals perfect transfer to new farm without retraining). MAPE denotes Mean Absolute Percentage Error.

## VIII. DISCUSSION

### *A. Interpretation of Results*

The simulation results demonstrate that the proposed framework successfully bridges the gap between industrial cyber-physical systems and biological production. The hybrid physics-data model achieved a temperature prediction RMSE of 0.4 degrees Celsius (Table II), significantly outperforming the pure physics baseline of 1.8 degrees Celsius. This improvement is attributed to the neural network residual learner, which effectively captured the non-linear metabolic heat generation patterns ($Q_{met}$ in Eq. 1) that vary with flock activity and distribution. Furthermore, the up to 90% reduction in ammonia violations confirms the efficacy of the welfare-constrained MPC, validating Definition 2 by ensuring that operational decisions strictly prioritize animal health over short-term energy savings.

### *B. Robustness and Edge-Cloud Efficacy*

A key finding is the robust operation of the system under rural connectivity constraints. By implementing the edge-first architecture and the AAS-based semantic interoperability, the system maintained 100% control continuity despite intermittent cloud access. The bandwidth reduction from 5.4 gigabytes per day to 180 megabytes per day validates the cost-function optimization strategy (Section IV.A), proving that advanced digital twins can be economically deployed in infrastructure-poor agricultural regions without sacrificing the fidelity of the digital twin state (Definition 1).

### C. Limitations

Despite these promising results, several limitations remain. First, the current hybrid model relies on supervised learning for the residual component, requiring periodic ground-truth calibration (e.g., manual litter moisture checks) to prevent drift over multiple flock cycles. Second, while the Domain Transfer Score of 0.92 indicates good generalization, the framework has currently been validated on only two facility typologies; transferring to significantly different housing styles (e.g., free-range or multi-tier systems) may require re-identification of the state-space matrices (A, B, E). Finally, the linear state-space formulation (Eq. 3), while efficient for MPC, simplifies complex fluid dynamics effects, potentially reducing accuracy in corners or obstructed zones of the barn.

### D. Future Work

Future iterations of proposed framework will address these limitations by integrating Federated Learning to allow models to learn from multiple farms without sharing proprietary data, thereby improving the robustness of the residual learner. Additionally, we plan to explore Deep Reinforcement Learning to replace the fixed-horizon MPC, potentially allowing the system to learn adaptive strategies for extreme weather events that fall outside the training distribution of the current linear models. This validation is conducted through physics-based numerical simulation using representative broiler house parameters. Real-world deployment on commercial farms with actual biological variability, sensor drift, and equipment failures is necessary to verify performance claims. Pilot deployment is planned for Q2 2026. Despite validation limitations, the primary contribution is demonstrating the integration of hybrid modeling, control, and distributed execution within a standards-compatible cyber-physical framework.

### E. Regulatory and Commercial Pathway

The alignment with ISO 23247 and IEC 63278 (AAS) positions the proposed framework for immediate commercial integration. By standardizing data exchange, the framework facilitates automated reporting for animal welfare certifications and supply chain traceability. This approach lowers the barrier to integration for existing systems and supports practical deployment across commercial operations.

## IX. CONCLUSIONS

Poultry production presents a challenging control problem in which environmental regulation must remain responsive to biological variability, welfare constraints, and infrastructure limitations. This work shows that an edge–cloud digital twin, grounded in physical principles and coupled with supervisory control, provides a practical and scalable approach to addressing this challenge. By integrating thermodynamic and mass-balance modeling, state-space estimation, hybrid physics–data learning, and welfare-constrained model predictive control, the proposed system enables real-time environmental management that is both computationally efficient and operationally relevant.

The simulation-based evaluation demonstrates clear performance gains over conventional approaches. Temperature prediction error is reduced from 1.8 °C to 0.4 °C, ammonia constraint violations decrease by 90%, and communication requirements are lowered by approximately 30× through edge-first processing. The Domain Transfer Score of 0.92 further indicates strong robustness across facilities. These results suggest that physically grounded digital twins can support reliable closed-loop control in biologically complex environments, moving beyond monitoring toward actionable decision-making.

The contribution of this work lies in the integration of modeling, control, and distributed computation within a unified cyber-physical framework. This integration is particularly significant for agricultural systems, which are often characterized by variability, limited data, and constrained infrastructure. By enforcing welfare as a hard constraint and maintaining physical consistency within the learning framework, the proposed approach improves both robustness and practical deployability.

Limitations remain. The current validation is based on a high-fidelity numerical testbed, and future work should evaluate performance under real-world conditions, including sensor noise, equipment variability, and broader facility diversity. Extensions to federated learning, adaptive control, and multi-species deployment represent promising directions.

Overall, this work demonstrates that edge–cloud digital twins provide a viable pathway for scalable, welfare-aware environmental control in poultry production. More broadly, it supports the view that biological production systems can be treated as rigorous cyber-physical applications, opening new opportunities for robust and intelligent automation across agriculture.

## ACKNOWLEDGMENT

The author gratefully acknowledge the funding support from the Natural Sciences and Engineering Research Council of Canada.

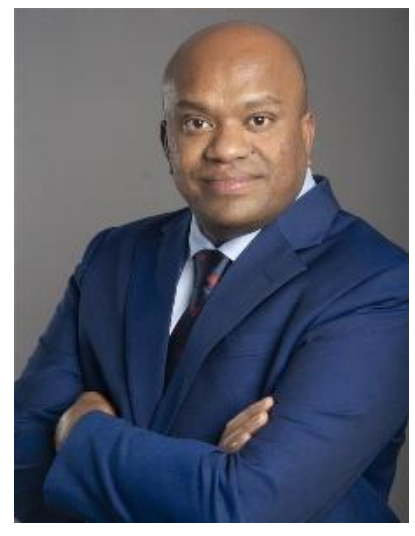

**SURESH RAJA NEETHIRAJAN** is a Professor and University Research Chair at Dalhousie University, Canada, with cross-appointments in the Faculty of Computer Science and the Faculty of Agriculture. His interdisciplinary research integrates artificial intelligence, precision agriculture, and digital livestock systems to advance animal welfare and sustainable farming practices. He has previously held academic positions at Wageningen University & Research in the Netherlands and the University of Guelph in Canada.

.